\pdfoutput=1
\documentclass{article}
\usepackage[table]{xcolor}
\usepackage{spconf,amsmath,graphicx}
\usepackage{pdfpages}
\usepackage{booktabs}
\usepackage{multirow}
\usepackage{arydshln}
\usepackage{pifont}
\usepackage{url}
\usepackage[numbers,sort,compress]{natbib} 
\definecolor{aliceblue}{rgb}{0.94, 0.97, 1.0}

\def\prepara{{\vspace{5pt}}}

\title{VoxSRC 2020: The second VoxCeleb Speaker Recognition Challenge}

\name{\em Arsha Nagrani$^1$, Joon Son Chung$^{1,2}$, Jaesung Huh$^{1}$, Andrew Brown$^{1}$, Ernesto Coto$^1$, Weidi Xie$^1$, \\ \em Mitchell McLaren$^3$, Douglas A Reynolds$^4$ and Andrew Zisserman$^1$}

\address{$^1$Visual Geometry Group, Department of Engineering Science, University of Oxford, UK\\
$^2$Naver Corporation, South Korea \\
$^3$Speech Technology and Research Laboratory, SRI International, Menlo Park, CA, USA \\
$^4$MIT Lincoln Laboratory, Lexington, MA, USA \\
\small{\url{http://www.robots.ox.ac.uk/\~vgg/data/voxceleb/competition2020.html}}}

\begin{document}
\maketitle
\begin{abstract}
We held the second installment of the VoxCeleb Speaker Recognition Challenge in conjunction with Interspeech 2020. The goal of this challenge was to assess how well current speaker recognition technology is able to diarise and recognize speakers in unconstrained or `in the wild' data. It consisted of: (i) a publicly available speaker recognition and diarisation dataset from YouTube videos together with ground truth annotation and standardised evaluation software;  and  (ii)  a  virtual public challenge and workshop held at Interspeech 2020.
This paper outlines the challenge, and describes the baselines, methods used, and results. We conclude with a discussion of the progress over the first installment of the challenge. 
\end{abstract}

\section{Introduction}
\label{sec:intro}
In 2019 we introduced the VoxCeleb Speaker Recognition Challenge~\cite{chung2019voxsrc} (VoxSRC), a new series of speaker recognition challenges that are intended to be hosted annually. The primary goals of VoxSRC are to: (i) explore and promote new research in speaker recognition ‘in the wild’; (ii) measure and calibrate the performance of the current state of technology through public evaluation tools; and (iii) provide open-source data freely accessible to all in the research community. 

While speech technologies have developed rapidly during the last few decades (with a large focus on automatic speech recognition and speaker verification), speaker recognition and diarisation under noisy and unconstrained conditions
are still extremely challenging. Applications of speaker
recognition are many and varied, ranging from authentication
in high-security systems and forensic tests, to high fidelity search of persons in large corpora of speech data. For such systems to be deployed in the real world, it is crucial that they work under unconstrained conditions, with noisy, varied and sometimes very short and fleeting speech segments. 



After a successful challenge and workshop in 2019, there was a constructive discussion on the limitations of VoxSRC2019 and ideas for extensions that would improve the quality of the challenge (Sec.6 of~\cite{chung2019voxsrc}). These were (i) the use of more metrics than simply EER, (ii) the addition of new tasks, and (iii) given the large number of submissions leading to satuation on the VoxSRC19 test set, a more challenging test set. 

We are pleased to note that VoxSRC2020 successfully incorporated all these three suggestions. The single biggest change for the challenge this year was the addition of a brand new task -- {\em speaker diarisation}, for which we also released a new dataset called~\textit{VoxConverse}~\cite{Chung20}. For speaker verification, an additional {\em self-supervised} track was added, and new metrics were introduced for both tasks (namely DCF for speaker verification, and JER and DER for speaker diarisation). Finally, we created an extremely challenging test set for speaker verification by incorporating out of domain data from movie material~\cite{brown2020playing,bain2020condensed}. 

In this paper, we describe the details of the evaluation task, the datasets provided, the challenge evaluation results and subsequent discussion. Further details can be found at the challenge website\footnote{\url{http://www.robots.ox.ac.uk/~vgg/data/voxceleb/competition.html}}. 

\section{Task Description} 
There were two tasks in this challenge, \textit{speaker verification} and \textit{speaker diarisation}. \textit{Speaker verification} is the task of determining whether a given pair of speech utterances are from the same speaker or not, while \textit{speaker diarisation} aims to break up multi-speaker audio into homogeneous single speaker segments, effectively solving ‘who spoke when’. Within the task of speaker verification we had three different tracks, each constraining the data allowed for training models, though with a common test and evaluation metrics. 

\subsection{Tracks} \label{sec:tracks}
The challenge consisted of the following four tracks:
\begin{enumerate}
    \item Speaker Verification -- Closed
    \item Speaker Verification -- Open
    \item Speaker Verification -- Self-supervised (Closed)
    \item Speaker diarisation -- Open
\end{enumerate}
 The first two tracks were identical to those in VoxSRC2019~\cite{chung2019voxsrc} (see also Sec.~\ref{sec:data}). 
 

Track 3 and 4 were new in VoxSRC2020. Inspired by recent successes in self-supervised learning~\cite{stafylakis2019self,lee2020momentum,nagrani2020disentangled,huh2020augmentation}, we introduced Track 3, where participants \textit{could not} use any speaker labels during training, however they were allowed to use the visual modality (faces) as well from the videos. 
We also introduced a speaker diarisation track (Track 4), as a new task this year. 
The open and closed training conditions refer to the training data allowed, and are described in Sec.~\ref{sec:data}.

\subsection{Data}\label{sec:data}
The VoxCeleb datasets were still the primary datasets for the speaker recognition (tracks 1--3). For VoxSRC2020 we also used a new set of speaker recognition segments from movie material, called \textbf{VoxMovies}~\cite{brown2020playing}, to create more challenging validation and test sets. 
For speaker diarisation (track 4), we introduced a new diarisation dataset from YouTube called \textbf{VoxConverse}~\cite{Chung20}. 
\subsubsection{Speaker Verification -- Track 1,2 and 3}
The VoxCeleb datasets~\cite{nagrani2020voxceleb,Chung18a,Nagrani17} consist of speech segments from unconstrained YouTube videos for several thousand individuals, and were created using an automatic pipeline.
For a full description of the pipeline and an overview of the datasets, see~\cite{nagrani2020voxceleb}. 


\prepara\noindent\textbf{Train set (Closed and Open Conditions):} The closed training condition required that participants train only on the VoxCeleb2 dev dataset~\cite{Chung18a}, which contains 1,092,009 utterances from 5,994 speakers. For the open training condition, participants could use the VoxCeleb datasets and any other data, except for the challenge's \textit{test} data. 

\prepara\noindent\textbf{Val and Test sets:} 
We provided a challenging validation set to participants to 
examine the performance of their models before uploading results to the evaluation server, in addition to the actual test set which was released a month before the challenge results were due. Unlike the validation set, the test set was \textit{blind}, i.e.\ the speech segments were released but with no annotations. The test data was released strictly for reporting of results alone, participants were not allowed to use this data in any way to train or tune systems.

The validation dataset consisted of trial pairs of speech from the identities in the VoxCeleb1 dataset, while the test set consisted of disjoint identities not present in either VoxCeleb1 or VoxCeleb2. Each trial pair consisted of two single-speaker audio segments, of variable length. 

In order to make a challenging validation and test set, we obtained some out-of-domain data for the same identities for which we had YouTube interview data for. These more challenging audio segments were sourced from the VoxMovies dataset~\cite{brown2020playing}, which contains speech segments from movie clips~\cite{bain2020condensed}. The VoxCeleb sourced segments, while valuable, are collected entirely from interviews on YouTube, and are limited in terms of linguistic content, emotion and background noise. On the other hand, the VoxMovies segments are sourced from an entirely different domain \textit{i.e.} movies, and contain speech covering many different emotions, accents, and varied background noise for the same identities. This out of domain data offers a significant challenge to state of the art speaker recognition systems, as shown in~\cite{brown2020playing}. 
The statistics of the val and test sets can be found in Table \ref{tab:testdata_track123}. The val and test data were checked using a combination of automatic and manual techniques for any errors using the same procedure described in \cite{nagrani2020voxceleb}, and following an identical procedure to VoxSRC19~\cite{chung2019voxsrc}.

In accordance with the feedback from last year, the challenge did not have same-session trials (e.g.\ segments from the same interview) in the test and validation sets. 

\subsubsection{Speaker Diarisation - Track 4}
The VoxConverse~\cite{Chung20} dataset is an audio-visual speaker diarisation dataset which includes 526 videos from YouTube. These videos are mostly from debates, talk shows and news segments. It has mutli-speaker, variable-length audio segments, with some overlap, and with challenging background conditions. Inspired by other audio-visual dataset creation pipelines such as VoxCeleb~\cite{Nagrani17} or VGGSound~\cite{chen2020vggsound}, it is generated from an automatic audio-visual speaker diarisation method using active speaker detection~\cite{Chung16a}, audio-visual source separation~\cite{Afouras18} and speaker verification. Only audio files are used for this workshop. Please refer to \cite{Chung20} for a more detailed description. 

\prepara\noindent\textbf{Train set:} Since the diarisation track is in the open training condition, participants are allowed to use any public or internal datasets except for the \textit{test} data. We provide a dev set for training and validation consisting of 216 wav files covering 1,216 minutes. The average number of speakers is 4.5 and the average overlap percentage of speech per video is 3.8\%.

\prepara\noindent\textbf{Val and Test sets:} We encouraged participants to use the VoxConverse dev set to validate their models. The VoxConverse test set contains 310 wav files, which total ~56 hours. It is more challenging than the VoxConverse dev set since both the average number of speakers and video duration is higher and the proportion of the audio track that is speech is lower. Details for both sets are described in Table 2 of~\cite{Chung20}.

\begin{table}[]
    \centering
    
    \begin{tabular}{ccccc}
    \toprule
         & \textbf{\# Pairs} & \textbf{\# Utter.} & \textbf{Segment length (s)}  \\ \midrule
         val  & 263,486 & 140,185 & 2.05/8.18/144.92\\
         test & 1,695,248 & 118,439 & 2.04/5.02/81.04\\
         \bottomrule
    \end{tabular}
    \caption{\small{ Statistics of the Speaker Verification val and test sets (Tracks 1--3). \textbf{\# Pairs} refers to the number of evaluation trial pairs, whereas \textbf{\# Utter.} refers to the total number of unique speech segments in the test set. Segment lengths are reported as min/mean/max.}}
    \label{tab:testdata_track123}
\end{table}

\begin{table*}[]
    \renewcommand\arraystretch{1.1}
    \centering
    
    \begin{tabular}{cccccc}
    \toprule
         \textbf{Track} & \textbf{Rank} & \textbf{Team Name} & \textbf{Organization} & \textbf{minDCF} & \textbf{EER}  \\ 
          \midrule
        \rowcolor{aliceblue} \multirow{4}{*}{1}  & - & Baseline & Provided & 0.477 & 7.68 \\
                  & 3 & ntorgashov~\cite{ntorgashov} & ID R\&D Inc., New York, USA & 0.203 & 3.82 \\ 
         & 2 & xx205~\cite{xiang2020xx205} & AISpeech Ltd, China & 0.196 & 3.81\\ 
        &1 & JTBD~\cite{thienpondt2020idlab} & IDLab, Ghent University, Belgium& 0.177 & 3.73\\ 
         
         \hline
        \rowcolor{aliceblue}\multirow{4}{*}{2}  & - & Baseline & Provided & 0.477 & 7.68 \\
         & 3 & DKU-DukeECE~\cite{wang2020dku} & Duke Kunshan University, China \& Duke University, USA & 0.205 & 3.88 \\          

         & 2 & xx205~\cite{xiang2020xx205} & AISpeech Ltd, China & 0.194 & 3.80\\ 
        & 1 & JTBD~\cite{thienpondt2020idlab} & IDLab, Ghent University, Belgium& 0.174 & 3.58\\ 

         \midrule
        \rowcolor{aliceblue}\multirow{4}{*}{3}  & - & Baseline & Provided & 0.877 & 19.07 \\
         &3 & umair.khan~\cite{khan2020upc} & TALP Research Center, UPC, Spain & 0.751 & 14.71\\
         &2 & DKU-DukeECE~\cite{wang2020dku} & Duke Kunshan University, China \& Duke University, USA & 0.598 & 12.42\\ 
  &1 & JTBD~\cite{thienpondt2020idlab} & IDLab, Ghent University, Belgium& 0.345 & 7.21\\ 
         \bottomrule
    \end{tabular}
    
    \caption{\small{Winners for the speaker verification tracks (Tracks 1,2 and 3). For both metrics, a lower score
is better.}}
    \label{tab:results_verification}
\end{table*}



\begin{table*}[]
    \renewcommand\arraystretch{1.1}
    \centering
    \begin{tabular}{ccccc}
    \toprule
         \textbf{Rank} & \textbf{Team Name} & \textbf{Organization} & \textbf{DER} & \textbf{JER}  \\ \midrule
\rowcolor{aliceblue}- & Baseline & Provided & 21.75 & 51.89 \\
         3 & DKU-DukeECE~\cite{wang2020dku} & Duke Kunshan University, China \& Duke University, USA & 9.79 & 20.67\\ 
         2 & landini~\cite{landini2020analysis} & Brno University of Technology, Czechia & 8.12 & 18.35\\ 
1 & mandalorian~\cite{xiao2020microsoft} & Microsoft Inc., USA& 6.23 & 21.52\\ 
         \bottomrule
    \end{tabular}
    \caption{\small{Winners for the speaker diarisation track (Track 4). For both metrics, a lower score
is better. }}
    \label{tab:results_diarsiation}
\end{table*}

\section{Challenge Mechanics} 
\subsection{Evaluation metrics}
We released a validation toolkit\footnote{\url{https://github.com/a-nagrani/VoxSRC2020}} for both speaker verification and speaker diarisation. Participants were encouraged to evaluate their models using this public code on the validation set of each track.

\prepara\noindent\textbf{Speaker verification. }
For the speaker verification tracks (track 1-3), we displayed two metrics, Equal Error Rate (EER) and minimum Detection Cost Function (minDCF). EER is a popular metric for evaluating the performance of speaker verification. It is used to determine the threshold value for a system when its false acceptance rate (FAR) and false rejection rate (FRR) are equal. minDCF ($C_{DET}$) can be computed as:
\begin{equation}
    C_{DET} = C_{miss} \times P_{miss} \times P_{tar} +C_{fa} \times P_{fa} \times (1 - P_{tar}) 
\label{eqn:dcf}
\end{equation}

This is same as the primary metric of the NIST SRE 2018 evaluation~\cite{nist2018}.
We set $C_{miss} = C_{fa} = 1$ and $P_{tar}=0.05$ in our cost function. 

For track 1 and 2, the primary metric was minDCF and final ranking was determined by this score alone. For track 3, the primary metric was EER. For both metrics, a lower score is better.

\prepara\noindent\textbf{Speaker diarisation.}
For track 4, we adopted two diarisation metrics, Diarisation Error Rate (DER) and Jaccard Error Rate (JER). DER is used as a primary evaluation metric in this track.

DER is a standard evaluation metric for speaker diarisation. It is the sum of speaker error, false alarm speech and missed speech. We applied a forgiveness collar of 0.25 sec, and overlapping speech was not ignored.

We also reported the Jaccard error rate (JER), a metric introduced for the DIHARD II challenge that is based on the Jaccard index. The Jaccard index is a similarity measure typically used to evaluate the output of image segmentation systems and is defined as the ratio between the intersection and union of two segmentations. To compute Jaccard error rate, an optimal mapping between reference and system speakers is determined and for each pair the Jaccard index of their segmentations is computed. The Jaccard error rate is then 1 minus the average of these scores. For more details please consult Section 3 of the Dihard Challenge Report~\cite{ryant2019second}.


\subsection{Baselines} 
We provided baselines (with open sourced code) for all tracks to help new participants get started. 
 For the fully-supervised speaker verification tracks (tracks 1 and 2), we provided a baseline consisting of a Fast ResNet-34 backbone trained on 40-dimensional mel spectrograms. The architecture and training procedures are described in detail in~\cite{chung2020defence}. The baseline achieved a minDCF of 0.477 and an EER of 7.68\% on the test set.   

For the self-supervised track, we trained a baseline model with a Fast ResNet-34 backbone, using contrastive learning and data augmentation (additive noise and room impulse response). Several similar techniques have already been introduced~\cite{inoue2020semi, huh2020augmentation}. This model achieved a minDCF of 0.877 and an EER of 19.07\% on the test set. 

For track 4, we used the baseline system from the second DIHARD challenge~\cite{ryant2019second}. This baseline is adopted from the JHU submission to the first DIHARD challenge~\cite{sell2018diarization}, which exploited standard clustering-based speaker diarisation. Speaker embeddings are extracted with a sliding window approach, followed by probabilistic linear discriminant analysis (PLDA) and aggolomerative hierachical clustering (AHC). We did not apply speech enhancement for pre-processing, resulting in a 21.75\% DER and 51.89\% JER on the challenge test set.

\subsection{Submission} 
Similar to the previous year, the challenge was hosted via CodaLab\footnote{\url{https://competitions.codalab.org/}} with two phases: ``Challenge workshop'' and ``Permanent''. Participants could only submit one submission per day in order to prevent overfitting on the challenge test set. Submission for the ``Challenge workshop" phase was available until 16th of October, 2020. For the last 48 hours before the final deadline, the leaderboard was made anonymous.

All teams that participated in the challenge were required to submit the challenge report describing their system by 23rd of October, 2020. The workshop was held on the 30th of October, 2020 in conjunction with Interspeech 2020.

\section{Methods and Results} 
272 submissions were made across all four tracks of the challenge. The top three performances for each track are shown in Tables~\ref{tab:results_verification} and ~\ref{tab:results_diarsiation}. Please refer to the challenge website for more details of results.

\prepara\noindent\textbf{Speaker verification.}
 Interestingly, the winners of all three speaker verification tracks were the same~\cite{thienpondt2020idlab}.
 
 For the fully-supervised tracks, the winner explored 6 variants of ECAPA-TDNN systems~\cite{desplanques2020ecapa} and 4 variants of the ResNet34~\cite{he2016deep} architecture. For the open track, the winner used the VoxCeleb1 dev set as well as additional speech data from the \textit{train-other-500} train set of the LibriSpeech dataset~\cite{panayotov2015librispeech} and a subset of the DeepMine corpus~\cite{zeinali2018deepmine} (babble, noise). Several data augmentation techniques were adopted for both the open and closed track -- including additional noise samples from the MUSAN corpus~\cite{snyder2015musan}, reverberation with RIR filters ~\cite{ko2017study} and the SoX and FFmpeg libraries for adjusting tempo and compression of speech. Large-margin fine-tuning was also applied with an AAM-softmax layer~\cite{deng2020sub} and quality-aware score calibration, both of which led to an improvement of 3\% and 8\% in terms of EER on the VoxSRC-20 test set, respectively. The final submission scored 0.177 minDCF in track 1 and 0.174 in track 2. The second place for both track 1 and 2~\cite{xiang2020xx205} used a fusion of ResNext~\cite{xie2017aggregated}, Res2Net~\cite{gao2019res2net} and a dual path network~\cite{chen2017dual} for the speaker network. They additionally explored the wide range of margin values of AAM-softmax, the dimension of output vector and the network size. Score normalization was used to boost the performance. The second place achieved a minDCF of 0.196 in track 1 and 0.194 in track 2.
We note that similarly to last year, the gap between the winning methods of Track 1 and 2 is not large (0.177 vs 0.174 min DCF), despite the fact that Track 2 (open condition) allows additional training data.
 
 In the self-supervised track, both the first and second place used a similar training framework. In both cases they (1) trained the network using contrastive learning, (2) generated pseudo-labels based on the model from the first stage, and (3) trained the network in a supervised way using these pseudo-labels. The first place team~\cite{thienpondt2020idlab} exploited Momentum Contrast (MoCo)~\cite{he2020momentum} for the first stage, followed by iterative clustering using both efficient mini-batch k-means and Agglomerative Hierarchical Clustering (AHC) to make pseudo-speaker labels. A large ECAPA-TDNN was then trained with these labels using a sub-center AAM-softmax~\cite{Deng2020SubcenterAB} layer. This was a similar technique to that used by the winners~\cite{thienpondt2020idlab} in their fully-supervised track submissions. The second place exploited a contrastive learning framework similar to~\cite{chen2020simple, falcon2020framework}, employed k-means with an extra purification step for pseudo-labels, and then trained the network for classification using a cross-entropy loss at the final stage. The first and the second place achieved an EER of 7.21\% and 12.42\% on the challenge test set, respectively.
 
 \prepara\noindent\textbf{Speaker diarisation.}
 Although this was the first appearance of a speaker diarisation track in a VoxSRC challenge, 43 submissions from 17 different teams were made. The performances of the three highest scoring teams are shown in Table~\ref{tab:results_diarsiation}.
 
 The winner~\cite{xiao2020microsoft} of this track exploited various novel techniques in their system. The audio input was first processed by a conformer-based~\cite{gulati2020conformer} continuous speech separation (CSS) technique, resulting in two separated channels. The individual channels were then fed into the Res2Net-based~\cite{gao2019res2net} speaker embedding extractor, which was trained with AM-Softmax~\cite{wang2018additive} loss. This was followed by Agglomerative Hierachical Clustering (AHC) with leakage filtering. Moreover, the outputs from multiple systems are fused by a modified version of the voting-based algorithm DOVER~\cite{stolcke2019dover}. The winner achieved 6.23\% DER on our test set.
 
 The second place~\cite{landini2020analysis} team adopted an LSTM-based speech enhancement technique, a ResNet152-based speaker embedding extractor and Variational Bayes Hidden Markov Model (VB-HMM) clustering. They also applied global speaker embedding re-clustering and a LSTM-based overlapping speech detector~\cite{bredin2020pyannote} trained with the AMI corpus~\cite{carletta2007unleashing} for post-processing. The resulting system achieved second place in terms of DER (8.12\%) and first place in terms of JER (18.35\%) on our test set.

 
\section{Workshop}
Due to COVID-19 and in line with Interspeech 2020, the VoxSRC 2020 workshop was held entirely virtually as a Zoom webinar. Once again the workshop was free of cost for anybody to attend. The number of attendees peaked at over 150 during the event, with a constant attendance of over 100 for the duration of the workshop. The workshop consisted of two keynote presentations: the first, from Dr Daniel Garcia-Romero gave a detailed summary on the recent history of speaker verification methods, titled ``X-vectors: Neural Speech Embeddings for Speaker Recognition''; while the second, from Professor Shinji Watanabe discussed methods for speaker diarisation, titled ``Tackling Multispeaker Conversation Processing based on Speaker diarisation and Multispeaker Speech Recognition''. This was in line with the introduction of a diarisation challenge track (track 4). Additionally there were announcements of the winners of each challenge track, and short presentations from the winners where they gave an overview of their methods. After each presentation, the speakers answered questions live from attendees. Over 50 live questions were asked to keynote speakers and challenge winners. All slides and recorded videos from the workshop are available at \url{http://www.robots.ox.ac.uk/~vgg/data/voxceleb/interspeech2020.html}. The workshop was kindly sponsored by Naver Corporation.

\section{Related Challenges}
Track 1,2 and 3 are focused on speaker recognition, which has been explored by the NIST-SRE (Speaker Recognition Evaluation) series~\cite{nist2018, sadjadi20172016, sadjadi20202019}, held since 1996 to measure state-of-the-art speaker recognition systems. Researchers from both academia and industry are encouraged to participate in NIST, however unlike NIST, all training data for VoxSRC is released publicly to the research community, even for those not participating in the challenge. Other challenges on speaker verification focus on noisy conditions~\cite{nandwana2019voices} or the far-field condition~\cite{qin2020interspeech}. 

Track 4 is complementary to several existing audio speaker diarisation challenges. The DIHARD challenges~\cite{ryant2018first,ryant2019second} are potentially the most popular. They evaluate state-of-the-art systems on extreme, ``hard'' conditions. Both the dev and test sets cover various background conditions, such as audiobooks, broadcast interviews, and restaurants. The third installment of the challenge~\cite{ryant2020third} will be concluded in early 2021.
Unlike VoxSRC, the challenge does not provide explicit training data, and hence any public or private data can be used for training models. Additionally, the DIHARD challenge applies \textit{no} forgiveness collar during evaluation and also has two separate diarisation tracks, one with oracle VAD and another with system VAD. 
Another popular challenge is the CHIME-6 challenge, where participants perform both speaker diarisation and speech recognition for multi-speaker conversations held in kitchen, dining and living room areas. The challenge data was made using binaural microphones and 4-channel microphone arrays, and the number of participants is fixed for each session. More details are provided at ~\cite{watanabe2020chime}.

\section{Discussion} \label{sec:discussion}
 \begin{table}[]
    \renewcommand\arraystretch{1.1}
    \centering
    \begin{tabular}{lcc}
    \toprule
         \textbf{Method} & \textbf{2019 test} & \textbf{2020 test}    \\ \midrule
         VoxSRC2019 winner~\cite{zeinali2019but} & 1.42 & - \\
         \midrule
          JTBD (VoxSRC2020) ~\cite{thienpondt2020idlab} & 0.80 &  3.73\\  
          xx205 (VoxSRC2020)~\cite{xiang2020xx205} & 0.75 & 3.81 \\
         \bottomrule
    \end{tabular}
    \caption{\small{Comparison of methods (\% EER) on the 2019 and the 2020 test sets, demonstrating that the 2020 test set is more challenging. We also compare the VoxSRC2019 winning submission and the top-2 submissions from VoxSRC2020 on the 2019 test set, showing the large performance improvement in a year. All results are shown on the closed track (Track 1). For \% EER shown, lower is better. }}
    \label{tab:results_comparison}
\end{table}
The workshop had wider attendance this year, potentially due to the virtual format. Additionally, all talks were pre-recorded and made accessible on the website, providing future access. While we hope future workshops will be in-person, to encourage open access, we will endeavor to record and livestream presentations during future workshops. 
Participation in the challenge was lowest for the self-supervised track, potentially because this is still a new area for speaker verification. We also note that the best performance for the self-supervised track (0.345 minDCF), is still far behind the full supervised tracks (0.177 minDCF) on the same test set. We also note that all methods used \textit{audio} only, with the visual modality not being utilised at all. This year the self-supervised track was closed (participants could only train on the VoxCeleb2 dev set), but in future years we may introduce additional tracks to determine if self-supervised methods can outperform fully supervised ones given sufficient training data (following the trend in computer vision~\cite{chen2020simple,he2020momentum} and other areas). 

While the VoxSRC2020 test set was larger that the VoxSRC2019 test set, with more challenging audio samples included from movie material, the VoxSRC2019 test set was still included in its entirety as a subset of the test set this year, allowing us to easily measure the performance of all submissions made this year on the 2019 test set alone. Table~\ref{tab:results_comparison} shows the performance of the top-2 submissions from VoxSRC2020 on both the 2019 and the 2020 test sets (bottom two rows), demonstrating that the 2020 test set is far more challenging. We also compare performance on the 2019 test set with that of the winner of VoxSRC2019. The top-2 winners of the challenge this year significantly outperformed 2019's winner, demonstrating the vast improvement in speaker verification performance over one year. Somewhat surprisingly, unlike the results shown in Table~\ref{tab:results_verification}, team xx205's performance is better than team JTBD's performance on the 2019 test set.

\subsection*{Acknowledgements}
This work is funded by the EPSRC Programme
Grant Seebibyte EP/M013774/1. Arsha Nagrani is funded by a Google PhD Fellowship. Andrew Brown is funded by an EPSRC DTA Studentship. Jaesung Huh is funded by a Global Korea Scholarship. This material is based upon work supported by the Air Force Research Laboratory under Air Force Contract No. FA8702-15-D-0001.  Any opinions, findings and conclusions or recommendations expressed in this material are those of the author(s) and do not necessarily reflect the views of the US Department of Defense. 
We would like to thank Daniel Garcia-Romero for his comments, Max Bain for his help with processing the Condensed Movies Dataset, and Triantafyllous Afouras and Abhishek Dutta for help with creating VoxConverse.

We also thank Rajan from Elancer and his team, \url{http://elancerits.com/}, for their huge assistance with diarisation annotation for VoxConverse. 

\bibliographystyle{IEEEbib}
\bibliography{shortstrings,refs}


\end{document}